\documentstyle[seceq,supplement,epsf,wrapft,mbf]{ptptex}

\newcommand{\beq}{\begin{equation}}
\newcommand{\eeq}{\end{equation}}
\newcommand{\beqa}{\begin{eqnarray}}
\newcommand{\eeqa}{\end{eqnarray}}
\newcommand{\ba}{\begin{array}}
\newcommand{\ea}{\end{array}}

\newcommand{\etal}{{\em et al} }

\pubinfo{~~~{\bf Preprint CAMTP/99-7}}  
\nofigureboxrule
\notypesetlogo  
\preprintnumber[3cm]{
CAMTP/99-6}

\begin{document}
\
\title{
Accuracy of the WKB approximation:
the case of \\general quartic potential
}

\author{
Marko {\sc Vrani\v car}\footnote{E-mail address: mark.vranicar@uni-mb.si}
and Marko {\sc Robnik} \footnote{E-mail address: robnik@uni-mb.si}
}

\inst{
CAMTP Center for Applied Mathematics and Theoretical Physics,
\\University of Maribor, Krekova 2, SI-2000 Maribor, Slovenia
}

\recdate{
}

\abst{
We analyse the accuracy of the approximate WKB quantization for the case
of general one--dimensional quartic potential. In particular, we are interested
in the validity of semiclassically predicted energy eigenvalues when
approaching
the limit $E\rightarrow \infty$, and in the accuracy of low lying energy levels
below the potential barrier in the case of generally asymmetric double--well
quartic potential. In the latter case, using the standard WKB 
quantization an unnatural  localization
of eigenstates  due to the negligence of tunneling is implied and thus the
validity of semiclassics is uncertain. In all computations the higher order
corrections to the leading semiclassical approximation are included using the
complex contour integration technique. We show that these corrections can
improve accuracy of semiclassical approximation greatly {\em by many orders of 
magnitude}.
}

\maketitle

\section{Introduction}

The semiclassical approximation is a very useful approximative tool in quantum
mechanics, not only because it often yields analytic expressions but also
because it helps us to understand the correspondence between quantum features
and classical mechanics, that is particularly important in the quantum chaos.

Certain rather naive thoughts about semiclassical approximation have been established in the
past. Namely, it has been  believed that the semiclassical quantization gives sufficiently
accurate energy values of quantum eigenstates, when $\hbar$ is sufficiently
small, in comparison to some action that stands with it, and in the case of
high energies as well. 
Both limits imply that the de Broglie wavelength must be sufficiently
small, but they are nevertheless {\em not} equivalent unless the underlying 
Hamiltonian has the scaling property.
But the crucial criterion for assessing the accuracy of 
semiclassically predicted energy values is the comparison of the absolute value of the error
of semiclassical energies with the mean energy level spacing (Prosen and
Robnik\cite{pr}). The fine structure, the statistical and other properties of 
energy spectra cannot be reliably resolved unless the error is small enough compared to
the mean energy level spacing.
 
The usual semiclassical techniques like EBK quantization (Maslov\cite{maslov}) 
and
Gutz\-wi\-ller trace formula are just the first (leading) terms of 
a certain semiclassical expansion of
quantum quantities in the $\hbar$ power series. 
Typically these are only asymptotic series, but occasionally can nevertheless 
be convergent, as a surprise (see below).
The absolute error of 
energies, 
predicted by those two methods is typically of order $\hbar^2$, while the mean
energy level spacing scales as $\hbar^f$, where $f$ is the number of degrees
of freedom (dimension of the
system's coordinate space). Thus, the two methods seem to be applicable only in
systems whose dimensionality is less than three. At a fixed value of $\hbar$ and in the
limit of high energies $E\rightarrow \infty$, the situation is even worse. The
error of the leading order semiclassical approximation, measured in the units
of the mean energy level spacing, can diverge or is bounded from below as $E$
goes to $\infty$  even in
cases of one and two dimensional systems (Prosen and
Robnik\cite{pr},  Robnik and Salasnich\cite{rs1,rs2}).
Nevertheless, for some specific systems, all of them exactly solvable, the WKB 
series can be convergent and its sum equal to the exact result. 
See ref.\cite{rs1,rs2,bender,rr1,rr2,ss}.

In this paper we analyse a one-dimensional general quartic potential, for
which we are able to calculate the systematic higher order correction in the
$\hbar$ expansion. Firstly we examine the simple case of pure  quartic 
potential, also called homogeneous quartic oscillator, which
is a scaling system, and thus we expect the $\hbar\rightarrow 0$ limit, to be 
equivalent to the $E\rightarrow \infty$ limit, so that the error of semiclassical energies
in units of the mean energy level spacing should converge to zero as energy 
increases.
Secondly, we tried to enrich our understanding of global behaviour of the
semiclassical approximation by examining its validity in the case of potential
with more than one minimum, like asymmetric double-well quartic potential.
A standard semiclassical quantization condition, obtained by requiring  that
the complex wavefunction as a function of the complexified coordinate 
be singlevalued, demands that
the change of the semiclassical phase around a circuit enclosing 
the two turning points (at the
boundaries of the classically allowed regions) is  an integer multiple of 
$2\pi$. We assume that neglecting the tunneling, as an exponentially 
small effect, this quantization condition can be also used in the 
cases 
where classically allowed regions are separated in the coordinate space with
the potential barrier.  In these cases motion in each well is meant to be
semiclassically quantized separately, and the error of energy values because of this unnatural 
localization is supposed to be exponentially small.
Strictly speaking, we actually do not know rigorously how to interpret this 
quantization condition, which is another important reason to study the implementation
of WKB methods and to test them in multi--well potentials like in our present work
(also see Robnik and Romanovski\cite{rr2}).
However, if one is interested explicitly in the fine structure of energy spectra, 
especially and in particular the exponentially small splitting (due to
tunneling) in {\em symmetric}
double well potentials, then the approach expounded in Robnik \etal\cite{rsv}
is the appropriate one. 

We show that the homogeneous and asymmetric nonhomogeneous quartic potential 
is indeed an example of a
potential where the semiclassical approximation works well in all energy ranges
that we were interested in. 
By adding higher order corrections, the accuracy of semiclassically predicted
energy values measured in units of the mean energy level spacing can be increased by many 
orders of magnitude.

\section{Homogeneous quartic potential  and semiclassical methods}

The Hamiltonian of the simple homogeneous quartic potential, that we consider in this section is
\beq
H=\frac{p^2}{2m}+Ax^4 =-\frac{\hbar^2}{2m}\frac{\partial\,^2}{\partial\,x^2}+Ax^4\,.
\eeq
Introducing the new scaled and dimensionless variables
\beq
\tilde{x}=x\left(\frac{mA}{\hbar^2}\right)^{1/6},~~~\tilde E=E\left(\frac{m^2}{\hbar^4A}\right)^{1/3},~~~\tilde
\hbar=1\,,
\eeq
the Hamiltonian can be rewritten in the simple form
\beq
\tilde H=-\frac{1}{2}\frac{\partial\,^2}{\partial\,\tilde x^2}+\tilde x^4\,. 
\eeq
In all further calculations only the scaled form of the Hamiltonian
and of the variables will be used, so that the tilde will be omitted
in the further writing. Note that the scaled energy is the only one 
system  parameter.

We study this system for two main purposes. Namely, we are interested in 
the asymptotic ($E\rightarrow \infty$) behaviour of standard 
semiclassical quantization formulas of finite order in the case of scaling system 
like this, and
further we want to compare the spectral properties of homogeneous quartic potential with the spectral properties of general 
asymmetric double well quartic potential which we also call non-trivial
potential. 
Again, in this comparison  we are mainly interested in the asymptotic
properties 
meaning high energy behaviour of the spectrum  of a  non-trivial potential whose 
geometry is asymptotically the same (as $|x|\rightarrow \infty$) as for the homogeneous
quartic potential. 
For these reasons we  use the standard WKB quantization procedure
rather than the implicit semiclassical formulas of Voros\cite{voros}.

To calculate the semiclassical spectrum of a Hamiltonian we observe that the 
wavefunction can always be
written in the form
\beq\label{hq4}
\psi (x) = \exp\left(\frac{i}{\hbar}\sigma (x)\right)\,,
\eeq
where  $\sigma (x)$ is the  complex phase that satisfies 
the Riccati differential equation (for $\sigma '$)
\beq\label{hq5}
\sigma{'}^2(x) + \frac{\hbar}{i} \sigma{''}(x) = 2m(E - V(x)) \, .
\eeq
The WKB approximation rests upon the  expansion of the phase in the $\hbar$ power series
\beq\label{hq6}
\sigma (x) = \sum_{k=0}^{\infty} \left(\frac{\hbar}{i}\right)^k \sigma_k(x) \, .
\eeq 
Substituting (\ref{hq6}) into (\ref{hq5}) and comparing like powers of $\hbar$ gives 
the recursion relation for $\sigma'_k(x)$ ($n>0$)
\beq\label{hq7}
\sigma{'}_0^2=2m(E-V(x)) \, , \;\;\;\; 
\sum_{k=0}^{n} \sigma{'}_k\sigma{'}_{n-k}
+ \sigma{''}_{n-1}= 0 \, .
\eeq

The quantization condition is obtained by requiring that the 
complex wavefunction (\ref{hq4}) is singlevalued as a function of the complexified 
coordinate, which means that
the change of phase, when going around  the classically allowed region 
between the two turning points $a$ ($V(a)=E$) and
 $b$ ($V(b)=E$), should be equal to an integer multiple of $2\pi\hbar$
\beq\label{hq8}
\oint d\sigma =\sum_{k=0}^{\infty} \left(\frac{\hbar}{i}\right)^{k}\oint d\sigma_k=2\pi\hbar n\,.
\eeq
The integer $n$ is the quantum number. The zeroth and the first order terms give the EBK or torus quantization
formula, where the first order term gives the Maslov corrections with Maslov index 2 for one dimensional system.
We observe that all the other odd terms in (\ref{hq8}) vanish when integrated along the closed contour around the pair
of turning points, because they are exact differentials which is a highly
plausible hypothesis yet to be proved rigorously
(Bender \etal\cite{bender},
Robnik and Romanovski\cite{rr2}) but it has been demonstrated to hold
for some low orders. We should stress that
the closed contour is the contour in the complex $x$ plane, since all integrals in (\ref{hq8}), except the zeroth order
term, diverge when integrated on an interval along real $x$-axes 
between the two turning points (Bender \etal\cite{bender}, 
Robnik \etal\cite{rsv}). However, there is a way of expressing these integrals
as a partial derivative w.r.t. the energy of a fundamental real integral
on the interval between the two turning points. 
Finally, taking this into account, the semiclassical quantization condition can be written as
\beq\label{hq9}
\sum_{k=0}^{\infty}\left(\frac{\hbar}{i}\right)^{2k}\oint d\sigma_{2k}=2\pi\hbar\left(n+\frac{1}{2}\right)\,,
\eeq
that is the sum over even-numbered terms only. The first three terms in 
(\ref{hq9}) are 
\beq\label{hq10}
\oint d\sigma_0=2\sqrt{2m}\int_a^b\sqrt{E-V(x)}\,dx,\;\;~~~~~~~~~V(a)=V(b)=E\,,\eeq
\beq\label{hq11}  
\oint d\sigma_2=\frac{1}{12\sqrt{2m}}\frac{\partial}{\partial\,E}\int_a^b\frac{V''(x)}{\sqrt{E-V(x)}}
\,dx\,,
\eeq 
\beq\label{hq12}  
\oint d\sigma_4=\frac{1}{576\sqrt{2m^3}}\left[
 \frac{7}{5}\frac{\partial\,^3}{\partial\,E^3}\int_a^b\frac{V{''}^2(x)}{\sqrt{E-V(x)}} \,dx-
\frac{\partial\,^2}{\partial\,E^2}\int_a^b\frac{V''''(x)}{\sqrt{E-V(x)}}\,dx\right]\,. 
\eeq  

A straightforward calculation of these terms for the case of the 
homogeneous quartic potential gives the semiclassical
quantization formula with corrections up to the fourth order
\[E^{3/4}\sum_{k=0}^{k=2}a_kE^{-3k/2}=\pi\left(n+\frac{1}{2}\right)\,;\]
\beq\label{hq13}
a_0=\frac{\Gamma \left(\frac{1}{4}\right)^2}{3\sqrt{\pi}}\,,~~~~~
a_1=-\frac{\pi\sqrt{\pi}}{4\Gamma \left(\frac{1}{4}\right)^2}\,,~~~~~
a_2=\frac{11\,\Gamma \left(\frac{1}{4}\right)^2}{6144\,\sqrt{\pi}}\,.\eeq

The exact spectrum is obtained by numerical diagonalization in 
QUADRUPLE PRECISION (32 valid digits) of the Hamiltonian matrix in the basis of
eigenstates of harmonic oscillator, whose Hamiltonian is 
\[H=\frac{1}{2} (\hat p^2+\omega^2x^2)\,.\]

The matrix elements in this basis are
\[
H_{n,m}=\left(\frac{\omega}{4}(2n+1)+\frac{3}{4\omega^2}(2n^2+2n+1)\right)\delta_{n,m}+
\]
\[ +\sqrt{(n+1)(n+2)}\left(-\frac{\omega}{4}+\frac{1}{2\omega^2}(2n+3)\right)\delta_{n+2,m}+\]
\beq +\sqrt{(n+1)(n+2)(n+3)(n+4)}\frac{1}{4\omega^2}\delta_{n+4,m}\,.\eeq 

\begin{figure}[t]
\epsfysize=7cm
\centerline{
\epsfbox{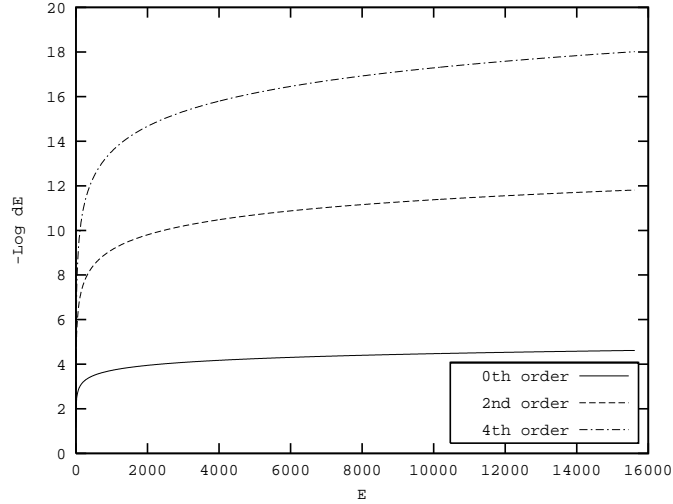}}
\caption{The  error 
of the 1100 lowest semiclassically predicted eigenenergies of the 
homogeneous quartic
potential (HQP). Negative decadic logarithm of the error in units of mean 
energy level spacing (-Log dE) is
plotted against energy E. The nearest level spacing is averaged over 7 nearest
 states to obtain the mean energy level spacing.}
\label{fig1}
\end{figure}

The results of our calculations are plotted in the 
figure \ref{fig1}, where the negative decadic logarithm of
the error of the semiclassical energies
obtained from the zeroth order, the second order and the fourth order 
semiclassical quantization formula in units of the mean
energy level spacing is plotted against energy. In table \ref{tab1} of the Appendix B we show the numerical values for some representative energy levels, showing the fast increase in accuracy with higher orders of the semiclassical approximation.

>From the plots we can deduce that  the error really goes towards zero
as we go to the high quantum numbers and/or energies. The accuracy of the semiclassics can be vastly enhanced, when
higher order corrections in $\hbar$ expansion are added to the EBK or torus quantization. Contrary to the
case of potential $U_0/\cos^2(\alpha x)$ previously studied 
(Prosen and Robnik\cite{pr},  Robnik and Salasnich\cite{rs1}), 
the homogeneous quartic potential is the case, where the $\hbar\rightarrow 0$ limit is equivalent to the
$E\rightarrow \infty$ limit as we predicted in the introduction, due 
to the scaling properties of our system.

\section{Asymmetric double-well quartic potential}

The Hamiltonian of the system that we consider in this section is generally written as
\beq\label{adwp1}
H=-\frac{\hbar^2}{2m}\frac{\partial\,^2}{\partial x^2}+Ax^4+Cx^3-2Bx^2\,.
\eeq
Again we introduce the scaled variables
\beq\label{adwp2}
\tilde x=x\sqrt{\frac{A}{B}},~~~\tilde
E=E\frac{A}{B^2},~~~\lambda=\frac{C}{\sqrt{AB}},~~~\hbar_{eff}=\hbar
\frac{A}{\sqrt{mB^3}}\,,
\eeq
and rewrite the Hamiltonian containing these variables
\beq\label{adwp3}
H=-\frac{\hbar_{eff}^2}{2}\frac{\partial\,^2}{\partial x^2}+x^4+\lambda x^3-2x^2\,.
\eeq

The notation with tilde has already been omitted in the equation (\ref{adwp3}).
The $\hbar_{eff}$ and $\lambda$ are two
additional parameters of the asymmetric double-well quartic potential. The latter is called the parameter of
asymmetry.  To check the accuracy of the semiclassical quantization for this type of the system, we perform
the  semiclassical quantization separately for states with energy below 
the potential barrier ($E<0$), and for
the states with energy above the potential barrier ($E>0$). 
\begin{wrapfigure}{r}{7cm}
\figurebox{60mm}{0cm}
\epsfxsize=6.0cm
\epsfbox{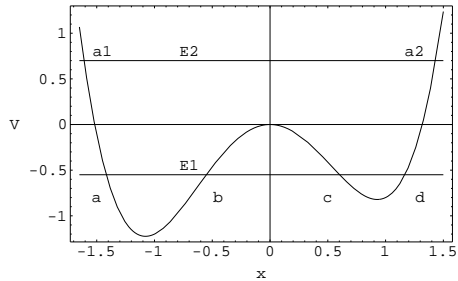}
\caption{A sketch of the asymmetric double-well quartic potential (ADWP)}
\end{wrapfigure}
Neglecting the tunneling, those states below the potential barrier, are semiclassically quantized using
the standard semiclassical procedure, described in the previous section, for both wells separately. 
We assume that
the error committed by this approximation and localization is only 
exponentially small, since the tunneling is
only exponentially small. The calculation of the integrals (\ref{hq10}), 
(\ref{hq11}) and (\ref{hq12}) that enter the
semiclassical quantization formula (\ref{hq9}) is much more complicated here than for the  system discussed in the
previous section.
The expressions obtained immediately after the integration on coordinate $x$
 in (\ref{hq11}) and (\ref{hq12}) are
very complex functions of the four roots of the momentum $p$, and energy.
These functions should be
further differentiated with respect to energy $E$. Please note, that the 
roots of
$p=\sqrt{2(E-V(x))}$ are also dependent on energy and in case of the fourth 
order correction formula (\ref{hq12}) a
three-fold differentiation with respect to $E$ should be done. The expression for the fourth
order corrections is thus too complex to be handled with our hardware equipment, regardless of
the fact that we do not need to know the explicit dependence on $E$ for the roots. The
derivative of a root $a$ with respect to the energy could be written in a form
$da/dE=(dV(x)/dx)^{-1}|_{x=a}$. 
In most of the manipulations we used the
{\em Mathematica} software. The calculation of expressions (\ref{hq10}) \
and (\ref{hq11}) are briefly sketched in the appendix A.

We have performed our calculations for three chosen values of 
$\hbar_{eff}$, namely 
at $\hbar_{eff}=0.01$,
$\hbar_{eff}=0.1$ and $\hbar_{eff}=1$, and at a fixed value of the parameter of asymmetry $\lambda=0.2$.
The shape of our asymmetric double well potential is shown in figure 2.
The negative decadic logarithm of the error of the energies calculated with the zeroth order and the second order
semiclassical quantization formula, in units of the mean energy level spacing,
is plotted in the figures \ref{fig3} and \ref{fig4}.

As we see from figures \ref{fig3}, \ref{fig4} and from the table \ref{tab2}
in the appendix B, the accuracy of semiclassical methods increases with energy like in the homogeneous case. The accuracy
is about 4 orders in magnitude or more greater when the second order corrections are included in the semiclassical
quantization formula. 

It is quite surprising how accurate the semiclassical quantization is even in the range of the energies below the
potential barrier whilst close to the top of the barrier it is not so good,
as expected. Of course we expect that the semiclassical methods are valid where the potential is similar to
the harmonic potential and our expectation seems to be proved now. 
As we see the accuracy decreases 
and the improvement of the semiclasically predicted energies
by considering the
higher order semiclassical corrections e.g. $\sigma_2$ 
gets smaller
as we approach the
top of the barrier. This is expected, since the exponentially small error caused by the negligence of tunneling
increases with energy, as we have shown in our previous work 
(Robnik \etal\cite{rsv}). 
Our results of this section are depicted in figures \ref{fig4}-\ref{fig6}.
Generally, the accuracy of
semiclassics increases as we decrease the value of $\hbar_{eff}$, what is of course expected and shown in figure \ref{fig6}.

Again, the exact spectrum has been calculated by numerical diagonalization 
in QUADRUPLE PRECISION of the Hamiltonian matrix in the basis
of eigenstates of the harmonic oscillator. Because the symmetry w.r.t.
parity operation is broken here, the matrix does
not decompose into even and odd part any more. The matrix elements of the matrix in the same harmonic oscillator 
\[H=\frac{1}{2} (\hat p^2+\omega^2x^2)\, ,\]
basis as before read as
\[ H_{n,m}=\left(\hbar_{eff}\left[\frac{\omega}{4}-\frac{1}{\omega}\right](2n+1)
+\frac{3\hbar^2_{eff}}{4\omega^2}(2n^2+2n+1)\right)\delta_{n,m}+ \]
\[+3\lambda\sqrt{\left[\frac{\hbar_{eff}}{2\omega}\right]^3}\sqrt{(n+1)^3}\delta_{n+1,m}+\]
\[
+\sqrt{(n+1)(n+2)}\left(-\hbar_{eff}\left[\frac{\omega}{4}+\frac{1}{\omega}\right]
+\frac{\hbar^2_{eff}}{2\omega^2}(2n+3)\right)\delta_{n+2,m}+\]
\[+\lambda\sqrt{\left[\frac{\hbar_{eff}}{2\omega}\right]^3}\sqrt{(n+1)(n+2)(n+3)}\delta_{n+3,m}+\]
\beq
+\sqrt{(n+1)(n+2)(n+3)(n+4)}\frac{\hbar^2_{eff}}{4\omega^2}\delta_{n+4,m}\,.\eeq 

\begin{figure}[h]
\parbox{7.8cm}{
\figurebox{7.8cm}{0cm}
\epsfxsize=7.5cm
\centerline{\epsfbox{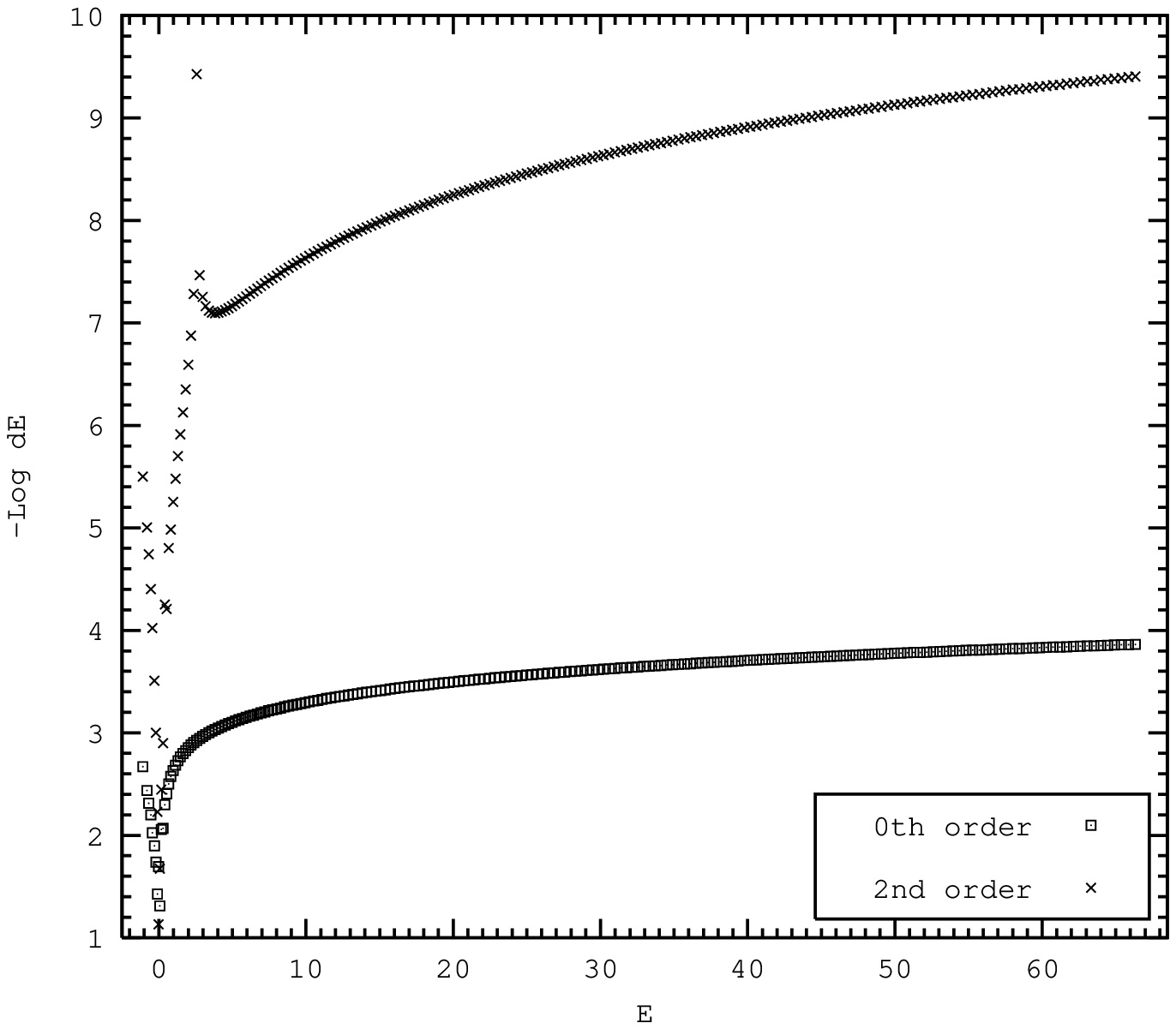}}}
\hspace{0mm} 
\parbox{6.2cm}{
\figurebox{6.2cm}{0cm}
\epsfxsize=6cm
\centerline{\epsfbox{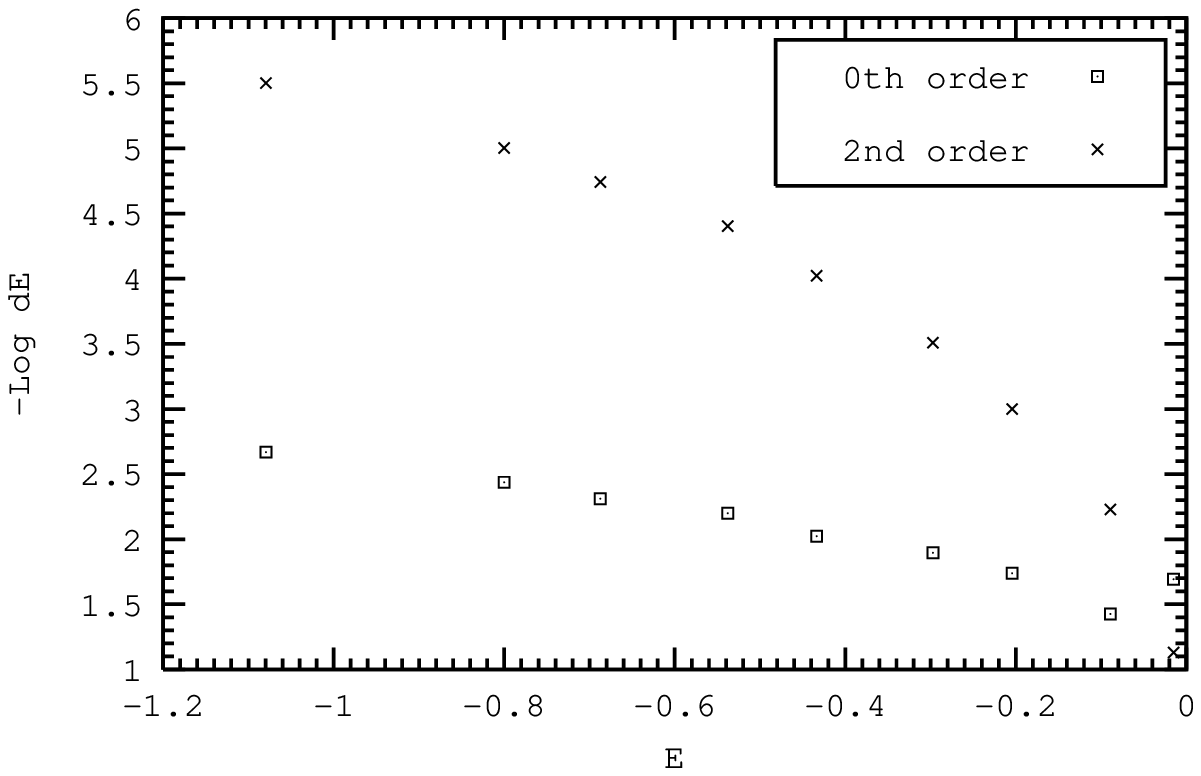}}}
\caption{Negative decadic logarithm of the error of the semiclassical energies in units of the mean energy
level spacing (-Log dE) {\em vs.} energy (E) of 200 lowest states (left), 
of which 9 lie below the barrier and are separately plotted in the right graph.
Asymmetric double-well quartic potential with $\hbar_{eff}=0.1$ and 
$\lambda=0.2$.}
\label{fig3}
\end{figure}

\begin{figure}[h]
\epsfxsize=7.5cm
\centerline{\epsfbox{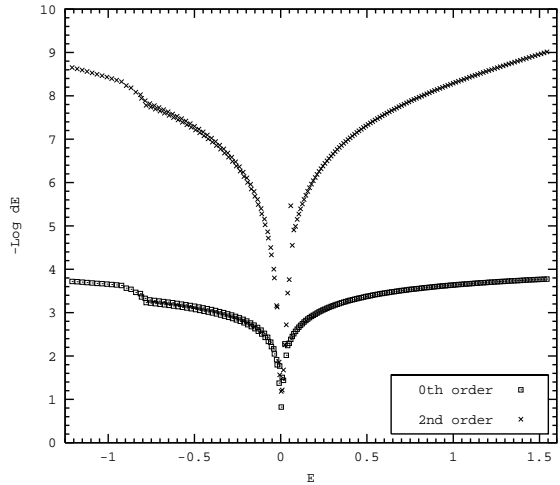}}
\caption{Negative decadic logarithm of the error of the
semiclassical energies in units of the mean energy level spacing (-Log dE) {\em vs.} energy (E) of 200 lowest
states, of which 86 lie below the barrier. Asymmetric double-well quartic potential with $\hbar_{eff}=0.01$ and $\lambda=0.2$.} 
\label{fig4}
\end{figure}

\begin{figure}[h]
\parbox{7cm}{
\figurebox{7cm}{0cm}
\epsfxsize=6.6cm
\centerline{\epsfbox{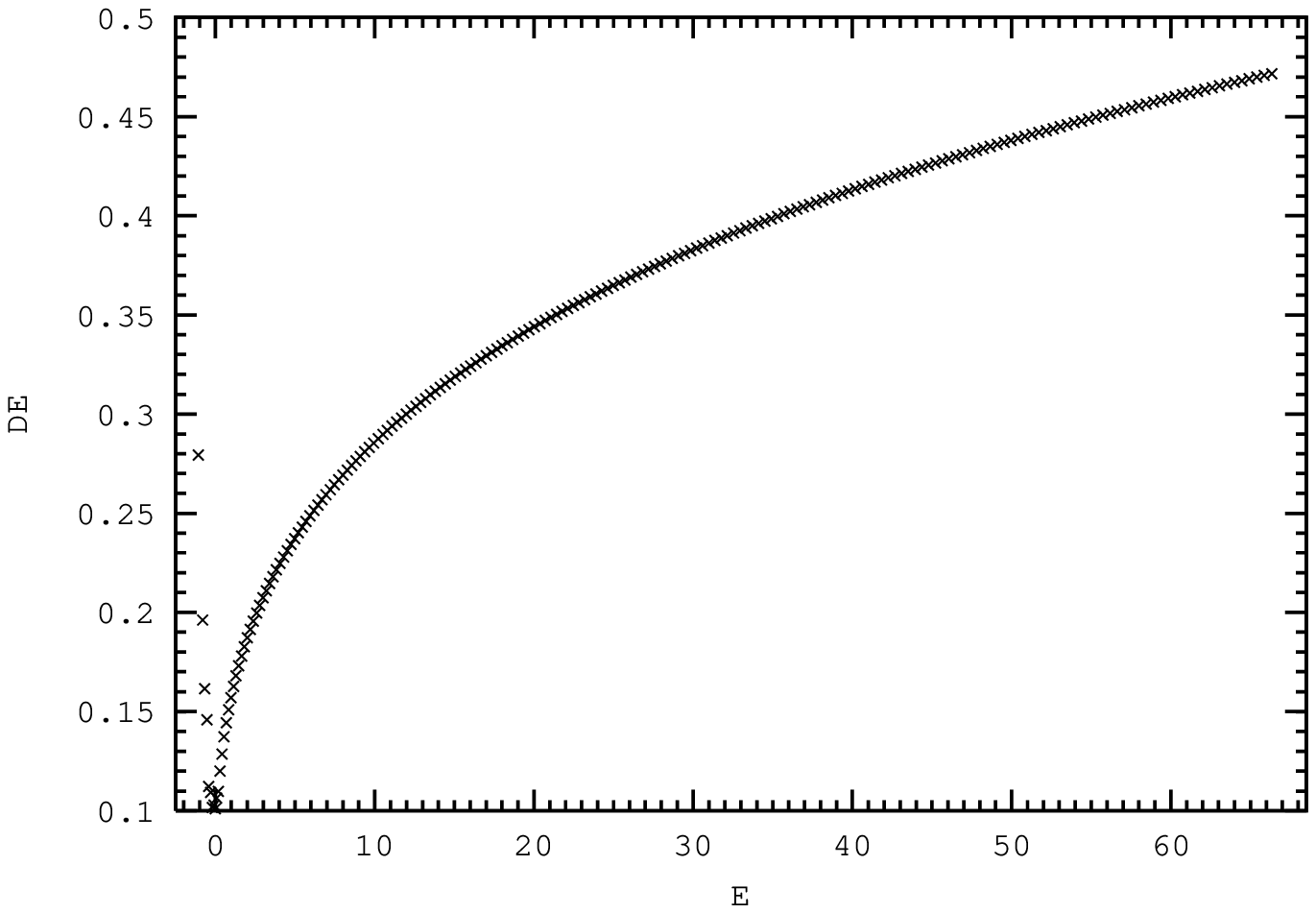}}}
\hspace{0mm} 
\parbox{7cm}{
\figurebox{7cm}{0cm}
\epsfxsize=6.6cm
\centerline{\epsfbox{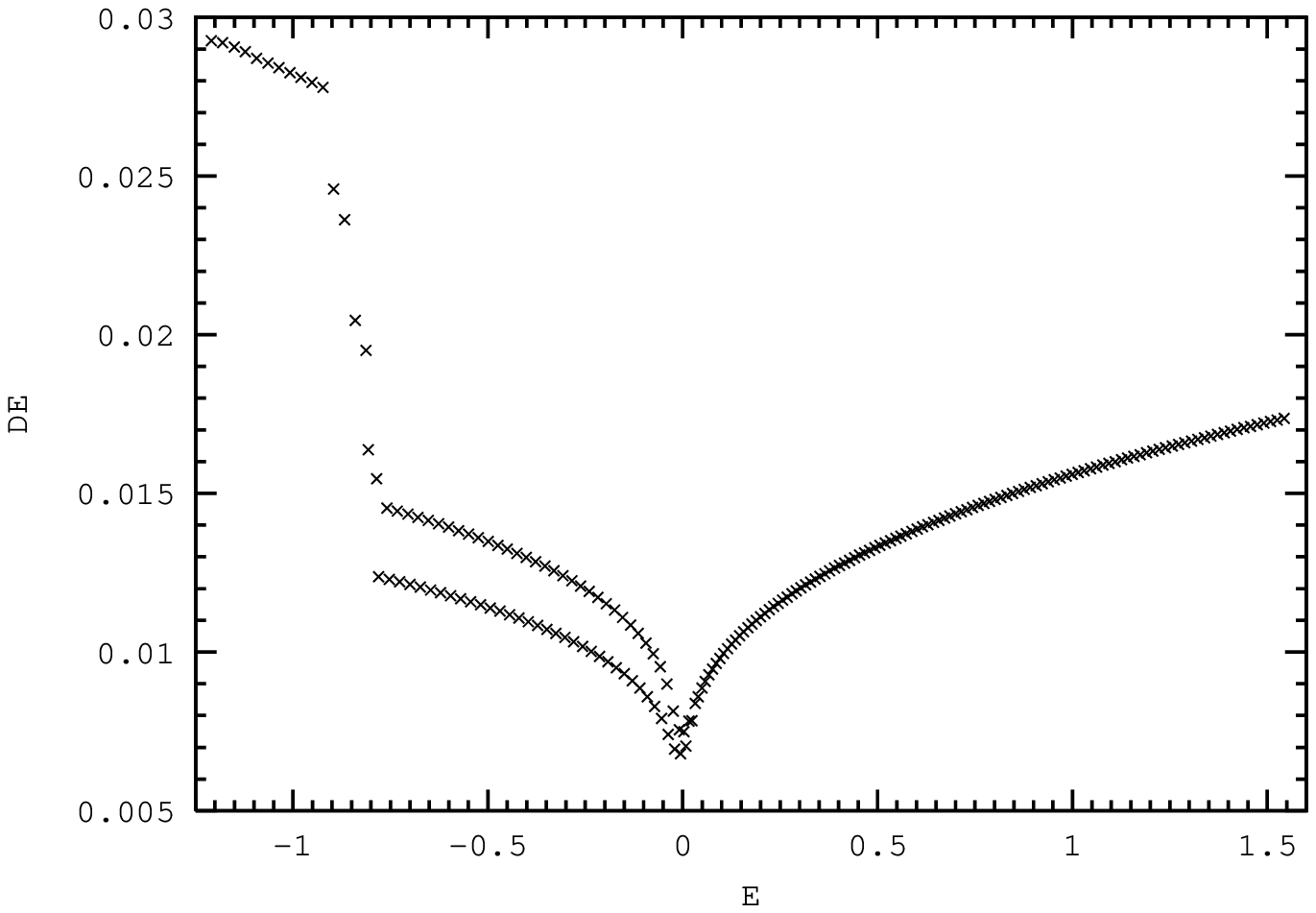}}}
\caption{The mean energy level spacing (DE),
averaged over 7 nearest states, {\em vs.} energy (E) of 200 lowest states. 
Asymmetric double-well quartic
potential with $\hbar_{eff}=0.1$ and $\lambda=0.2$ (left) and
$\hbar_{eff}=0.01$ and $\lambda=0.2$ (right).} 
\label{fig5}
\end{figure}

\begin{figure}[h]
\epsfxsize=6.6cm
\centerline{\epsfbox{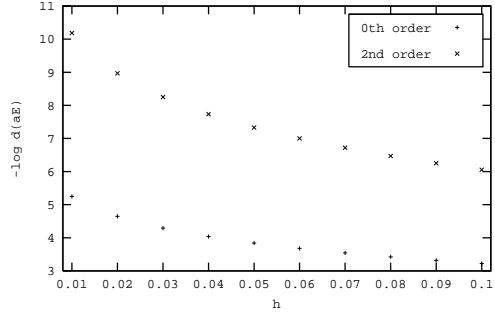}}
\caption{Negative decadic
logarithm of the absolute error of the semiclassical energies of the ground state
(-Log d(aE)) {\em vs.} $\hbar_{eff}$ (h). Asymmetric double-well quartic potential with
 $\lambda=0.2$.}
\label{fig6}
\end{figure}

\section{High energy correspondence between  the exact spectra of 
the homogeneous  and the general quartic potential }

During our work new questions about the spectral properties of the systems 
with non-trivial potential have arisen. By the word non-trivial potential we
mean e.g. the asymmetric double-well quartic potential and similar potentials with complicated structure in certain regions
of coordinate space  and with a simple asymptotic behaviour as $|x|\rightarrow \infty$. The question is, whether the
spectra of the homogeneous quartic potential and that of the asymmetric double-well potential with $\hbar=1$ 
converge pointwise or at least to a certain shift in energy and in the quantum number, as 
$E\rightarrow \infty$. 

\begin{figure}[h]
\parbox{7.8cm}{
\figurebox{7.8cm}{0cm}
\epsfxsize=7cm
\centerline{\epsfbox{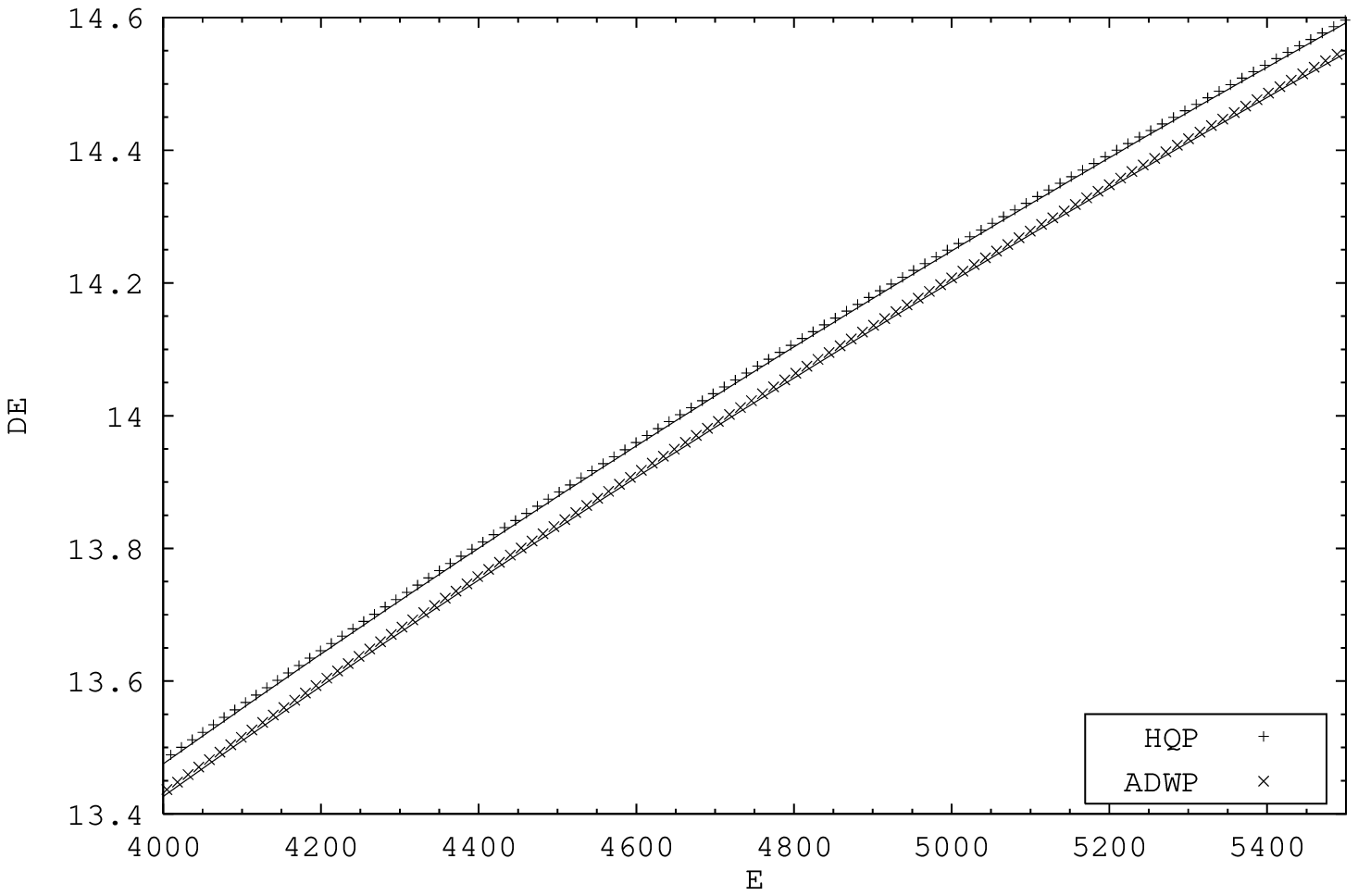}}}
\hspace{0mm} 
\parbox{6.2cm}{
\figurebox{6.0cm}{0cm}
\epsfxsize=6.0cm
\centerline{\epsfbox{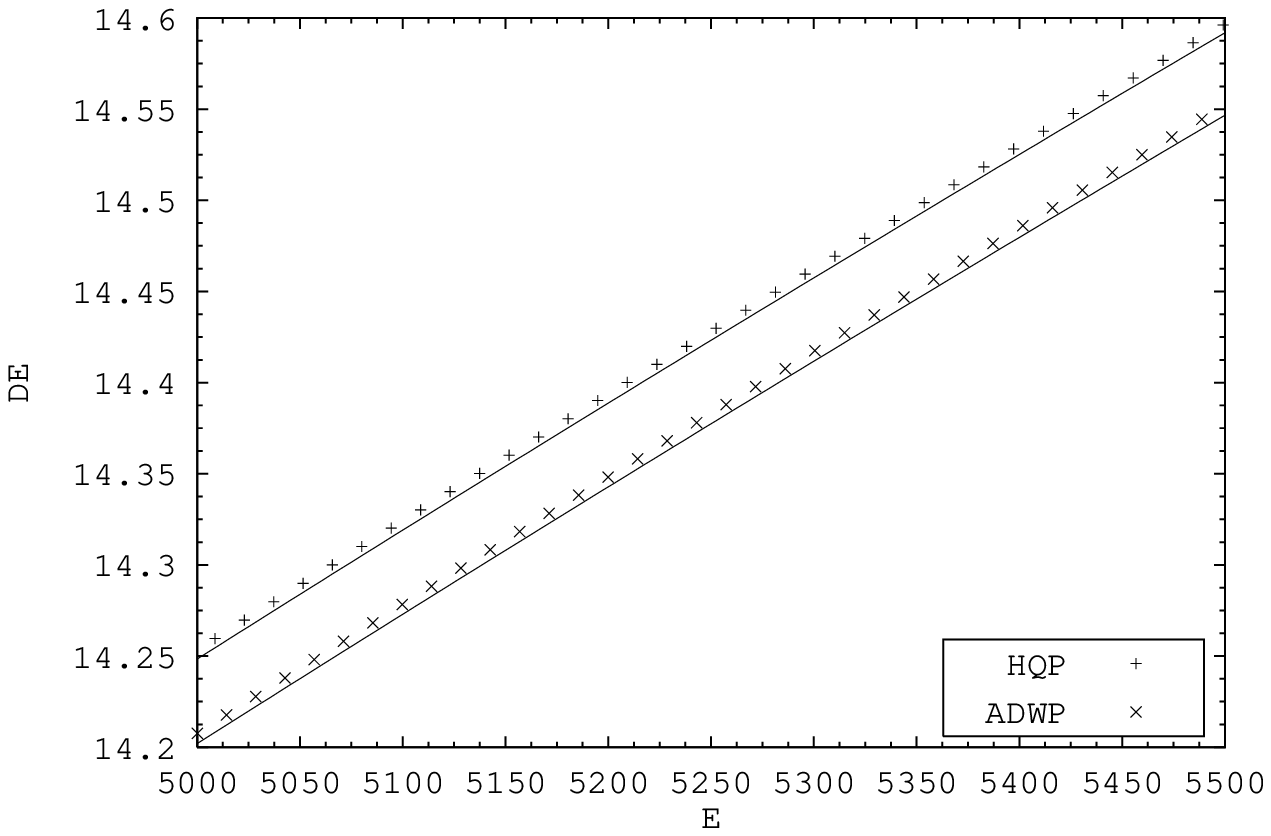}}}
\caption{Mean energy level spacing DE {\em vs.} energy E of the states between 400th and 500th state of HQP and ADWP ($\mu =1,~\lambda =0.2$) (left). 
The right figure shows an enlarged window of the left figure at the energy 
range $[5000,5500]$.
Solid line is the zeroth order semiclassical prediction
(Thomas-Fermi rule).}
\label{fig7}
\end{figure}

\begin{figure}[h]
\epsfxsize=7.5cm
\centerline{\epsfbox{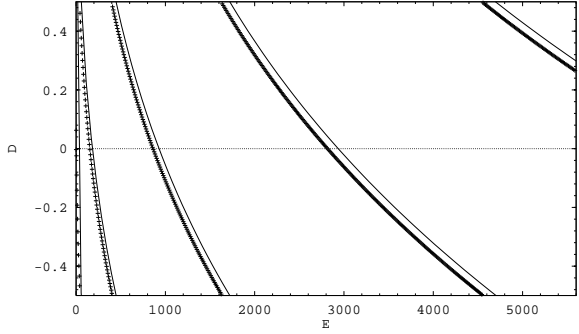}}
\caption{Energy difference of energetically nearest states of $\mu =0$ case 
(HQP) and $\mu =1,~\lambda=0.2$ case (ADWP)
{\em vs.} energy E of about 500 lowest states of HQP. 
The difference or discrepancy D is measured in units of 
mean energy level spacing of
the HQP. Solid line is the zeroth order semiclassical prediction.}
\label{fig8}
\end{figure}

The scaled  Hamiltonian of the general quartic potential can be written as
\[
H=\frac{p^2}{2}+x^4+\mu V_1(x)
\]
\beq V_1(x)=\lambda x^3-2x^2\,.\eeq
In the following figures we present the comparison of the spectrum for the case
 $\mu=0$ with the spectrum of the system
where $\mu=1,~\lambda=0.2$. We compare about 500 lowest states. In the 
figures \ref{fig7} the mean energy level spacing
against energy is plotted  for both systems. The discrepancy between the 
curves for the two systems is clearly seen. 
To any state of $\mu=0$ case we find the state of $\mu=1$ case which is energetically most close to it and in
figure \ref{fig8}  we plot
the discrepancy of those energies measured in the units of the mean energy level spacing of the $\mu =0$ case,  against the energy of $\mu=0$ states. We clearly see some cut-offs that occur
at some energy values, where the difference in quantum numbers of the energetically nearest  states of the two systems is 
enlarged by 1. 

To understand these features we calculate the zeroth order semiclassical approximation for the mean energy level spacing $DE$ (the Thomas-Fermi formula).
\beq\label{DE1}
 DE_{ADWP}=\frac{dE}{dn}=\pi\left[\int_{a_1}^{a_2}\frac{dx}{p(x)}\right]^{-1},~~~~
p=\sqrt{2(E-V(x))},~~~~p(a_1)=p(a_2)=0\, .
\eeq
For the case of the asymmetric double well potential (ADWP) the integral in
the expression above becomes
\beq\label{DE2}
 \int_{a_1}^{a_2}\frac{dx}{\sqrt{E-V(x)}}=\frac{2}{\sqrt{kq}}{\bf F}(m)
\eeq
\[ 
k^2=(a_1-b)^2+c^2,~~~~q^2=(a_2-b)^2+c^2,~~~~
m=\frac{kq-(a_1-b)(a_2-b)-c^2}{2kq}\, ,
\]
where $b$ and $\pm c$ are the real and the imaginary part of the other
two complex conjugate roots of $p(x)$ (see appendix A).
To calculate the mean energy level spacing for the ADWP case we expand
the 4 roots of $E-V(x)$ into the energy $E$ asymptotic series. 
The first few terms of those series are

\beqa\label{DE3}
a_{1,2}&=&\mp E^{1/4}-\frac{\lambda\mu}{4}\mp\left[\frac{3(\lambda\mu)^2}{32}+
\frac{\mu}{2}\right]E^{-1/4}-\lambda\mu\left[\frac{(\lambda\mu)^2}{32}+
\frac{\mu}{4}\right]E^{-1/2} \nonumber \\ 
&&\mp \frac{15(\lambda\mu)^4+160(\lambda\mu)^2\mu
+256\mu^2}{2048}E^{-3/4}+O\left(E^{-5/4}\right)\, ,
\eeqa
\beqa\label{DE4}
b\pm ic&=&-\frac{\lambda\mu}{4}+\lambda\mu\left[\frac{(\lambda\mu)^2}{32}+
\frac{\mu}{4}\right]E^{-1/2} 
+i\left\{E^{1/4}
-\left[\frac{3(\lambda\mu)^2}{32}+\frac{\mu}{2}\right]E^{-1/4}\right\}
\nonumber \\  
&&+i\left\{\frac{15(\lambda\mu)^4+160(\lambda\mu)^2\mu
+256\mu^2}{2048}E^{-3/4}\right\}
+O\left(E^{-5/4}\right)\, .
\eeqa
Now, we can do the same with the quantities $2/\sqrt{kq}$ and $m$
\beq\label{DE5}
\frac{2}{\sqrt{kq}}=\sqrt{2}E^{-1/4}-\frac{3(\lambda\mu)^4
+32(\lambda\mu)^2\mu+64\mu^2}{128\sqrt{2}}E^{-5/4}
+O\left(E^{-7/4}\right)\, ,
\eeq
\beq\label{DE6}
m=\frac{1}{2}+\left[\frac{3(\lambda\mu)^2}{32}+\frac{\mu}{2}\right]E^{-1/2}
+O\left(E^{-3/2}\right)\, .
\eeq
Bearing in mind that
\beq\label{DE7}
\frac{\partial {\bf F}(m)}{\partial m}=\frac{1}{2m(1-m)}
({\bf E}(m)-(1-m){\bf F}(m))
\eeq
we expand the elliptic integral ${\bf F}(m)$ into the asymptotic series
around $m=1/2$.
Rewriting 
\beq\label{DE8}
{\bf F}(1/2)=\frac{\Gamma(1/4)^2}{4\sqrt{\pi}}
\eeq
and
\beq\label{DE9}
{\bf E}(1/2)=\frac{\Gamma(1/4)^2}{8\sqrt{\pi}}\left(1+\frac{8\pi^2}
{\Gamma(1/4)^4}\right)\, ,
\eeq
finally we get
an asymptotic expression for the mean energy level spacing
\beq\label{DE10}
DE_{ADWP}=E^{1/4}\frac{4\sqrt{\pi}^3}{\Gamma(1/4)^2}\left[
1-E^{-1/2}\frac{\pi^2}{4\Gamma(1/4)^4}(16\mu+3(\lambda\mu)^2)\right]\, .
\eeq
It is easy to see that the factor outside the brackets is just
the mean energy level spacing of the homogeneous  quartic potential (HQP)
\beq\label{DE11}
DE_{HQP}=E^{1/4}\frac{4\sqrt{\pi}^3}{\Gamma(1/4)^2}\, .
\eeq
The zeroth order semiclassical predictions for the mean energy level spacing
(\ref{DE10}) and (\ref{DE11}) are also plotted in figures \ref{fig7}.
The result (\ref{DE10}) can be obtained also perturbatively by a systematic
perturbative treatment of the integral (\ref{DE1}).

Further we analyze the difference $D$ of energetically nearest 
states of HQP and
ADWP in units of mean energy level spacing of HQP. For this purpose
we expand the zeroth order semiclassical quantization condition for 
ADWP
\[
\int_{a_1}^{a_2}pdx 
=\frac{4\sqrt{2}}{3}\frac{1}{\sqrt{kq}}\left\{
\left[E-\frac{(\lambda\mu)\mu}{4}\frac{a_2k-a_1q}{k-q}
+\left(\mu+\frac{3(\lambda\mu)^2}{16}\right)\frac{a_2^2k-a_1^2q}
{k-q}\right]\,{\bf F}(m)+\right.
\]  
\beq\label{DE12}
+\left(\mu+\frac{3(\lambda\mu)^2}{16}\right)kq\,{\bf E}(m)+
\eeq
\[
+\left.\frac{3\lambda\mu}{8}\left(\mu+\frac{(\lambda\mu)^2}{8}\right)\frac{(a_2-a_1)(k+q)}{k-q}\,{\mbf \Pi}(n,m)\right\}=
\pi\left(M+\frac{1}{2}\right)\,;~~~~~n=-\frac{(k-q)^2}{4kq}
\]
(see  the end of the appendix A)
into the asymptotic energy series, valid at high energies. 
To do that we follow the similar procedure as in the case of derivation of the mean energy level spacing asymptotic expression (\ref{DE10}). Here we also
use the first order expansion of the three elliptic integrals for small arguments
$\epsilon$ and $\delta$
\[{\bf F}(1/2+\epsilon)={\bf F}(1/2)+(2{\bf E}(1/2)-{\bf F}(1/2))
\epsilon\,,
\]
\[{\bf E}(1/2+\epsilon)={\bf E}(1/2)+({\bf E}(1/2)-{\bf F}(1/2))
\epsilon\,,
\]
\[
{\mbf \Pi} (0+\delta,1/2+\epsilon)={\bf F}(1/2)+(2{\bf E}(1/2)-{\bf F}(1/2))\epsilon+2({\bf F}(1/2)-{\bf E}(1/2))\delta\,,
\]
in our case
\[
\epsilon =m-\frac{1}{2}=\left[\frac{3(\lambda\mu)^2}{32}
+\frac{\mu}{2}\right]E^{-1/2}
+O\left(E^{-3/2}\right),~~~~~
\delta=n=O\left(E^{-3/2}\right)
\]
and so we get an asymptotic expression of the condition (\ref{DE12}) 
\beq\label{DE13}
\pi\left(M+\frac{1}{2}\right)=E^{3/4}
\frac{\Gamma(1/4)^2}{3\sqrt{\pi}}
\left[1+E^{-1/2}\frac{3\pi^2}{4\Gamma(1/4)^4}
  (16\mu+3(\lambda\mu)^2)
\right]+O\left(E^{-1/4}\right)\,. \eeq 
Note, that
the leading term of the upper equation (\ref{DE13}) is precisely the
zeroth order semiclassical quantization condition for HQP 
(see (\ref{hq13}))
\beq\label{DE14}
\pi\left(N+\frac{1}{2}\right)=E^{3/4}\frac{\Gamma(1/4)^2}
{3\sqrt{\pi}}\,.
\eeq 
Since the $M$-th energy eigenvalue $E+\Delta E$  of ADWP is 
energetically nearest to the $N$-th eigenenergy $E$ of the 
HQP (by definition $\Delta E$ is smaller than 
one half of the mean energy level spacing), we expand the difference of equations (\ref{DE13}) and (\ref{DE14}) into the 
$\Delta E/E$ power series up to the first order and solve it 
for $\Delta E$. 
Finally we can write an approximate expression for the energy difference of the energetically closest states of ADWP and HQP at high energies
\beq\label{DE14a}
\Delta E=DE_{AQWP}\left[(M-N)-\frac{\sqrt{\pi}}{4\Gamma(1/4)^2}
(16\mu+3(\lambda\mu)^2)E^{1/4}\right]\, ,
\eeq
of which the leading term in the units of mean energy 
level spacing of HQP     
is
\beq\label{DE15}
D=\frac{\Delta E}{DE_{HQP}}\approx (M-N)-\frac{\sqrt{\pi}}{4\Gamma(1/4)^2}
(16\mu+3(\lambda\mu)^2)E^{1/4}\, .
\eeq
The above expression is also plotted in the figure \ref{fig8}
together with the exact results. 

According to the expression (\ref{DE15}) and also with respect to the exact results, plotted in figure \ref{fig8}  we can 
conclude that the two spectra do not converge pointwise and not  even converge
to a constant energy shift since the expression (\ref{DE15}) as a function of energy never reaches the constant value of $D$ at a fixed value of difference between the appropriate quantum numbers 
$N$ and $M$ of the homogeneous quartic potential and the asymmetric double well potential. 

The expression (\ref{DE10}) that fits the exact data quite well indicates that
the mean energy level spacing of the asymmetric double well 
potential 
converges to the mean energy level spacing of HQP as energy 
increases but it
converges as slow as $E^{-1/4}$.

\section{Conclusions}

According to our results we can conclude that the semiclassical approximation gives sufficiently good results in
the case of the quartic potential. As we see  the semiclassical energies in units of the mean energy level spacing
converge to the exact ones for the case of homogeneous quartic potential as $E\rightarrow \infty$. Due to this
scaling properties of the system we can talk about only one limit in this particular system 
where the semiclassics becomes exact, namely
the semiclassical limit which is equivalent to the $\hbar\rightarrow 0$ or $E\rightarrow \infty$
limit. The accuracy of the semiclassical energies at the high energy values is not spoiled even in the case of asymmetric
double-well quartic potential, which is not a scaling system. 
Even in the energy range below the potential
barrier the accuracy of the semiclassics is very good, and it is 
bounded with the exponential error due to the
negligence of tunneling. It is encouraging that the  higher order 
corrections in the $\hbar$ expansion of the
semiclassical quantization formula increase the accuracy by many orders of 
magnitude. Obviously, the corrections like
these are needed in two or more dimensional systems to prevent the error to diverge or to be bounded from below
as we approach $\hbar\rightarrow 0$ or $E\rightarrow \infty$ limit. 
Some recent related papers are Robnik and Salasnich\cite{rs1,rs2}, 
Robnik and Romanovski\cite{rr1,rr2}, Salasnich and Sattin\cite{ss}.
Unfortunately we are not able yet to derive
the systematic corrections of higher orders in $\hbar$ expansion to the EBK quantization in more than one 
dimensional systems. Nevertheless this work helps us to understand the global behaviour and limitations of
techniques based on semiclassical ideas. As the higher order corrections 
yield more accurate
results in one dimensional systems we might hope that similar ideas will help 
to improve the accuracy of the
semiclassics in many dimensional cases.

In section 4 we  discuss the asymptotic behaviour of exact 
spectra of nontrivial potentials. By the word
nontrivial we mean that the potential in certain range of coordinate possesses a complicated structure, but
asymptotically as $|x|\rightarrow \infty$ becomes similar to a simple 
(scaling) correspondent. We have a strong evidence, exact numerical and 
semiclassical, that
the spectra of both systems do not converge pointwise to the same energy values and not even to a constant energy shift as the energy
is increased. Even the convergence of the mean energy level spacing of the 
two systems is very slow when $E\rightarrow \infty$  (as slow as $E^{-1/4}$). 

\appendix
\section{
Calculation of the zeroth order and the second order terms in the 
semiclassical quantization formula}

In this section we show how to calculate integrals $\oint d\sigma_0$ and $\oint d\sigma_2$ for the case of
asymmetric double-well quartic potential. The integrals that enter the semiclassical quantization condition are
calculated  separately for the regions where the momentum $p=\sqrt{2(E-V(x))}$ has four and two real zeros.

The contour integrals around the pairs of turning points entering the zeroth and the second order semiclassical formula (\ref{hq10}) and (\ref{hq11}) 
respectively can be written as
\beq\label{start}
\oint \sqrt{X}\,dx=\frac{2}{3}\left[ E\oint \frac{dx}{\sqrt{X}}
-\frac{\lambda}{4}\oint \frac{x\,dx}{\sqrt{X}}
+\left(1+\frac{3\lambda^2}{16}\right)\oint \frac{x^2\,dx}{\sqrt{X}}\right]\, ,
\eeq
\beq\label{start1}
-\oint \frac{X''\,dx}{\sqrt{X}}=12\left[-\frac{1}{3}\oint \frac{dx}{\sqrt{X}}
+\frac{\lambda}{2}\oint \frac{x\,dx}{\sqrt{X}}
+\oint \frac{x^2\,dx}{\sqrt{X}}\right]\, ;\eeq
\[X=E-x^4-\lambda x^3+2 x^2=\frac{p^2}{2}\, .\]

All integrals that we need are
\[I_1=\oint \frac{dx}{\sqrt{X}}\, ,~~~I_2=\oint \frac{x\, dx}{\sqrt{X}}\, ,~~~
I_3=\oint \frac{x^2\, dx}{\sqrt{X}}\]
and will be calculated separately for the case
where  $X$ has four real roots and for the case where $X$ has two real roots.

\subsection{Momentum $p$ and $X$ have four real roots ($a<b<c<d$)}

The integrals are written for the case of contour integration around a pair of turning points $a$ and $b$. In case
of integration around $c$ and $d$ only substitution $a\longleftrightarrow c$, $b\longleftrightarrow d$ has to be
done.
\[X=E-x^4-\lambda x^3+2 x^2=(x-a)(b-x)(x-c)(x-d)\]

Firstly, we perform a  substitution
\[x=\frac{c-dy^2}{1-y^2}\]
in integrals $I_1,~I_2$ and $I_3$ and we get
\[I_1=\frac{4}{\sqrt{(d-a)(d-b)}}\int_{y_1}^{y_2}\frac{dy}{\sqrt{(y^2-y_1^2)(y_2^2-y^2)}}=\frac{4}{\sqrt{(d-a)(d-b)}}J_1;\]
\[ y_1=\sqrt{\frac{b-c}{b-d}},~~~~~y_2=\sqrt{\frac{a-c}{a-d}},\]

\[I_2=\frac{4}{\sqrt{(d-a)(d-b)}}\left[
d\int_{y_1}^{y_2}\frac{dy}{\sqrt{(y^2-y_1^2)(y_2^2-y^2)}}
-\right.\]
\[\left.
-(d-c)\int_{y_1}^{y_2}\frac{dy}{(1-y^2)\sqrt{(y^2-y_1^2)(y_2^2-y^2)}}\right]=\]
\[=\frac{4}{\sqrt{(d-a)(d-b)}}[dJ_1-(d-c)J_2],\]

\[I_3=\frac{4}{\sqrt{(d-a)(d-b)}}\left[
d^2\int_{y_1}^{y_2}\frac{dy}{\sqrt{(y^2-y_1^2)(y_2^2-y^2)}}-\right.\]
\[-\left. 2d(d-c)\int_{y_1}^{y_2}\frac{dy}{(1-y^2)
\sqrt{(y^2-y_1^2)(y_2^2-y^2)}}+ \right.\]
\[\left.+(d-c)^2\int_{y_1}^{y_2}\frac{dy}{(1-y^2)^2\sqrt{(y^2-y_1^2)(y_2^2-y^2)}}
\right]=\]
\[=\frac{4}{\sqrt{(d-a)(d-b)}}[d^2J_1-2d(d-c)J_2+(d-c)^2J_2^{(2)}]\, .\]
Factorizing the term $1/(1-y^2)^2$, $J_2^{(2)}$ can be rewritten in the form
\[J_2^{(2)}=\frac{1}{2}J_2+\frac{1}{4}S\, ,\]
\[S=\int_{y_1}^{y_2}\frac{dy}{(1-y)^2\sqrt{(y^2-y_1^2)(y_2^2-y^2)}}
+\int_{y_1}^{y_2}\frac{dy}{(1+y)^2\sqrt{(y^2-y_1^2)(y_2^2-y^2)}}\, .\]

To calculate $S$, we use the following relations
\[w^2=a_0y^4+4a_1y^3+6a_2y^2+4a_3y+a_4=\]
\[=b^{(c)}_0(y-c)^4+4b^{(c)}_1(y-c)^3
+6b^{(c)}_2(y-c)^2+4b^{(c)}_3(y-c)+b^{(c)}_4\]
and
\[
b^{(c)}_0\int_{y_1}^{y_2}
\frac{(y-c)^2\,dy}{w}
+2b^{(c)}_1\int_{y_1}^{y_2}
\frac{(y-c)\,dy}{w}-
2b^{(c)}_3\int_{y_1}^{y_2}\frac{dy}{(y-c)w}
=b^{(c)}_4\int_{y_1}^{y_2}
\frac{dy}{(y-c)^2w}\, ;
\]
\beq\label{a1}
w(y_1)=w(y_2)=0,~~~~~c\notin (y_1,y_2)
\eeq
(see Magnus \etal\cite{magnus}).

Inserting $w^2=(y^2-y_1^2)(y_2^2-y^2)$ and
evaluating the upper expression (\ref{a1}) for 
$c=1$ and $c=-1$ and subtracting what we get in both cases, we can express $S$ as
\[S=4\left[ \frac{b^{(1)}_0-2b^{(1)}_1}{2b^{(1)}_4}J_1+
\frac{b^{(1)}_3}{b^{(1)}_4}J_2+
\frac{b^{(1)}_0}{2b^{(1)}_4}J_3\right]\, .\]

\[b^{(\pm 1)}_0=-1,~~~b^{(\pm 1)}_1=\mp 1,~~~
b^{(\pm 1)}_2=\frac{y^2_1+y^2_2}{6}-1\,,
\]
\[
b^{(\pm 1)}_3=\pm \left(\frac{y^2_1+y^2_2}{2}-1\right),~~~
b^{(\pm 1)}_4=(y^2_1+y^2_2)-y^2_1y^2_2-1\,.
\]

Integrals 
\[J_1=\frac{1}{y_2}{\bf F}(m)\,,~~~~~
J_2=\frac{1}{y_2}{\bf F}(m)+\frac{y_1^2}{y_2(1-y_1^2)}{\mbf \Pi} (n,m)\,,~~~~~
J_3=y_2{\bf E}(m)\, ,
\]
\[
m=\frac{y^2_2-y_1^2}{y_2^2},~~~~~~
n=\frac{m}{1-y_1^2}
\]
can be found in (Gradshteyn \etal\cite{grad}) or (Abramowitz and 
Stegun\cite{abramowitz}).
Functions ${\bf F}(m)$, ${\bf E}(m)$ and ${\mbf \Pi} (k,m)$ are the complete elliptic
integrals of the first, the  second
and the third kind as defined by Abramowitz and Stegun.

Finally we can write
\[
\oint d\sigma_0=\frac{8\sqrt{2}}{3}\frac{1}{\sqrt{(d-b)(c-a)}} \]
\[\left\{\left[E-\frac{1}{2}\left(1+\frac{3\lambda^2}{16}\right)
(cd+ab)-\frac{3c\lambda}{4}\left(1+\frac{\lambda^2}{8}\right)\right]\,{\bf F}(m)+\right.\]
\[+\frac{1}{2}\left(1+\frac{3\lambda^2}{16}\right)(d-b)(c-a)\,{\bf E}(m)
+\]
\beq\label{a2}
+\left.\frac{3\lambda}{4}\left(1+\frac{\lambda^2}{8}\right)(c-b)\,
{\mbf \Pi} (n,m)\right\}\,;\eeq
\[m=\frac{(d-c)(b-a)}{(c-a)(d-b)},~~~n=\frac{b-a}{c-a}\,, \]
and
\beq\label{a3}
\oint
d\sigma_2=-\frac{1}{\sqrt{2}}\frac{\partial}{\partial\,E}\frac{1}{\sqrt{(d-b)(c-a)}}
\left\{\left(cd+ab+\frac{2}{3}\right)\,{\bf F}(m)
-(d-b)(c-a)\,{\bf E}(m)\right\}.\eeq

\subsection{ Momentum $p$ and $X$ have two real roots ($a_1<a_2$, $b\pm ci$)}

\[X=E-x^4-\lambda x^3+2 x^2=(x-a_1)(a_2-x)((x-b)^2+c^2)\]

 In this case the substitution 
\[
x=\frac{a_2+a_1\frac{q}{k}\tan^2\frac{\phi}{2}}{1
+\frac{q}{k}\tan^2\frac{\phi}{2}};\]
\[k^2=(a_1-b)^2+c^2,~~~~q^2=(a_2-b)^2+c^2\]
is performed in integrals $I_1,~I_2$ and $I_3$ and following the similar 
procedure
as in the case of 4 real roots of $X$, the integrals can be rewritten 
in the following form
\[I_1=\frac{4}{\sqrt{kq}}{\bf F}(m)\, ;\]
\[m=\frac{kq-(a_1-b)(a_2-b)-c^2}{2kq},~~~~n=-\frac{(k-q)^2}{4kq}\,,\]

\[I_2=\frac{4}{\sqrt{kq}}\left[\frac{a_2k-a_1q}{k-q}{\bf F}(m)
-\frac{(a_2-a_1)}{2}\frac{k+q}{k-q}{\mbf \Pi} (n,m)\right]\, ,\]

\[I_3=\frac{4}{\sqrt{kq}}\left[\frac{a^2_2k-a^2_1q}{k-q}{\bf F}(m)
+kq {\bf E}(m)
-(a_1+a_2+2b)
\frac{(a_2-a_1)}{4}\frac{k+q}{k-q}{\mbf \Pi} (n,m)\right]\, .\]

By using these expressions, $\oint d\sigma_0$ and $\oint d\sigma_2$
can be finally written as
\[ \oint
d\sigma_0=\frac{8\sqrt{2}}{3}\frac{1}{\sqrt{kq}}\left\{
\left[E-\frac{\lambda}{4}\frac{a_2k-a_1q}{k-q}
+\left(1+\frac{3\lambda^2}{16}\right)\frac{a_2^2k-a_1^2q}{k-q}\right]\,
{\bf F}(m)+\right.\]  
\[+\left(1+\frac{3\lambda^2}{16}\right)kq\,{\bf E}(m)+\]
\beq\label{a4}
+\left.\frac{3\lambda}{8}\left(1+\frac{\lambda^2}{8}\right)\frac{(a_2-a_1)(k+q)}{k-q}\,{\mbf \Pi}(n,m)\right\}\,;\eeq

\[k^2=(a_1-b)^2+c^2,~~~~q^2=(a_2-b)^2+c^2\, ,\]
\[m=\frac{kq-(a_1-b)(a_2-b)-c^2}{2kq},~~~~n=-\frac{(k-q)^2}{4kq}\,.\]
and 
\beq\label{a5}
 \oint
d\sigma_2=\frac{1}{\sqrt{2}}\frac{\partial}{\partial\,E}\frac{1}{\sqrt{kq}}
\left\{
\left(2\frac{a_2^2k-a_1^2q}{k-q}+\lambda\frac{a_2k-a_1q}{k-q}-\frac{2}{3}\right)\,{\bf F}(m)
+2kq\,{\bf E}(m)\right\}\, .\eeq

In the section 4 we use the generalized form of the asymmetric 
quartic potential
\[
V(x)=x^4+\mu (\lambda x^3-2x^2)\Rightarrow
X=E-x^4-\lambda\mu x^3+2\mu x^2=\frac{p^2}{2}\,.
\]
In this case the integral (\ref{start})
is slightly modified to
\beq\label{start2}
\oint \sqrt{X}\,dx=\frac{2}{3}\left[ E\oint \frac{dx}{\sqrt{X}}
-\mu\frac{\lambda\mu}{4}\oint \frac{x\,dx}{\sqrt{X}}
+\left(\mu+\frac{3(\lambda\mu)^2}{16}\right)\oint 
\frac{x^2\,dx}{\sqrt{X}}\right]\, ,
\eeq
yielding the $\mu$ dependent expression (\ref{DE12})
for $\sigma_0$, 
which is identical to (\ref{a4}) for $\mu=1$. Please note that
the parameters $k,~q,~m,~n$ as well as
the integrals 
$I_1$, $I_2$ and $I_3$ 
are expressed in terms of the four roots 
of $X$ and momentum and so is  their general dependence 
on $\mu$  already
considered.

\clearpage

\section{Some tables of results}
\begin{table}[h]
\caption{Some eigenenergies of the homogeneous quartic potential. By $n$ we denote the consecutive  quantum number of the state, while  $E_{Exact}$ is the exact energy value of the eigenstate, and  $E_{Semi}^{(0)}$, $E_{Semi}^{(2)}$, and
$E_{Semi}^{(4)}$  are the semiclassically predicted energy eigenvalues of 
the $n$th state, calculated with the zeroth order, second order and the 
fourth order semiclassical quantization formula.}
\begin{center}
\footnotesize
\begin{tabular}{lrrrr} \hline\hline
 & & & & \\[-3mm]
\footnotesize $n$ &\footnotesize $E_{Exact}$ &
\footnotesize $E_{Semi}^{(0)}$ & 
\footnotesize $E_{Semi}^{(2)}$  &
\footnotesize $E_{Semi}^{(4)}$   \\
 & & & & \\[-3mm] \hline
          0 &     0.6679862591       & 
 0.5462673250        & 
 0.6178440470      & 
 0.5994975436       \\

& 5577710827  & 7809440985&  9454695175 & 8186297455 \\

1 &     2.3936440164        & 
  2.3635614446         & 
  2.4003574561         & 
  2.3991273384         \\

&8230311603 &0057710124 &2199197003 & 3857621598\\

          2 &     4.6967953868          & 
  4.6705199316         & 
  4.6968569596         & 
  4.6965336443         \\
& 6364619622& 9480733314 & 9157646544 & 2176087907\\

          3 &     7.3357299952          & 
  7.3148026019         & 
  7.3358836919         & 
  7.3357508981         \\
&2709279884 & 0083816986 & 7629989381 & 3385710476\\

          4 &     10.2443084554         & 
  10.2265364336        & 
  10.2443782077         & 
  10.2443100572         \\
&3877107597 & 1101923867 & 8948408161 & 3952052722 \\

... & ... & ... & ... & ... \\

        100 &     643.1833913927         & 
  643.1811392054         & 
  643.1833914100        & 
  643.1833913927         \\
        
&8056430848 & 7609028834 &  9319827383& 8185826816\\

101 &     651.7305795719         & 
  651.7283422013        & 
  651.7305795888        & 
  651.7305795719        \\
        &  7519147803 &  2623003831 &  3668259301 &  7642701540 \\

103 &     668.9091219363         & 
  668.9069134812         & 
  668.9091219523        & 
  668.9091219363         \\
        & 1714901544  & 9348071280  &  2366319599 &  1827702401 \\

105 &     686.1986791466         & 
  686.1964986907         & 
  686.1986791618        & 
  686.1986791466         \\
        & 2251646322  & 3792741559  & 3254876433  &  2354809723 \\ 

... & ... & ... & ... & ... \\

495 &     5397.2204747691         & 
  5397.2196972873         & 
  5397.2204747693         & 
  5397.2204747691         \\
        & 2444635012  &  4414211246 &  7029195001 &  2444710621 \\ 

496 &     5411.7486521189         & 
  5411.7478756815         & 
  5411.7486521192         & 
  5411.7486521189        \\
        &  9592537080 &   1512620174&  4045276381 &  9592611981 \\ 

497 &     5426.2865864811         & 
  5426.2858110844        & 
  5426.2865864813         & 
  5426.2865864811        \\
        & 2223840485  & 4128681921  & 6545729016  & 2223914687  \\

500 &     5469.9588010681         & 
  5469.9580287730         & 
  5469.9588010683         & 
  5469.9588010681        \\
& 5121968046  & 5845185750  & 9057033639  &  5122040195 \\ 
... & ... & ... & ... & ... \\

1000 &     13774.2520020553         & 
  13774.2515153776        & 
  13774.2520020553         & 
  13774.2520020553        \\
       &  0377443322 &  5917115936 &  4151998141 &  0377446170 \\

1005 &     13866.1108150810        & 
  13866.1103300181         & 
  13866.1108150811         & 
  13866.1108150810         \\
&  7240132753 & 5261252080  & 0964842617  & 7240135535  \\

1008 &     13921.2992811341         & 
  13921.2987970336        & 
  13921.2992811341        & 
  13921.2992811341         \\
&   4444693588   &  5119763791 & 8139930100 &  4444696331\\        

1009 &     13939.7076086294        & 
  13939.7071248487         & 
  13939.7076086295         & 
  13939.7076086294        \\
& 9093649576  &  4694282718 &  2779132907 & 9093652307  \\       

1010 &     13958.1220154946         & 
  13958.1215320331        & 
  13958.1220154946         & 
  13958.1220154946        \\

& 3736333880 & 1516055405 & 7412099391 & 3736336599\\

\hline
\end{tabular}
\end{center}
\label{tab1}
\end{table}

\begin{table}[h]
\caption{Some eigenenergies of the asymmetric double--well
quartic potential with $\hbar_{eff}=1,~\lambda=0.2,~(\mu=1)$. By $n$ we denote the consecutive 
quantum number of the state, while 
 $E_{Exact}$ are the exact energy value of the $n$th eigenstate and
 $E_{Semi}^{(2)}$  are the semiclassically predicted 
energy eigenvalues of 
the $n$th state, calculated with the second order 
semiclassical quantization formula.}

\begin{center}
\footnotesize
\begin{tabular}{lrr} \hline\hline
 & &   \\[-3mm]
$n$ & $E_{Exact}$ & $E_{Semi}^{(2)}$ \\ 
 & &   \\[-3mm] \hline
          0 &   -0.17296458264338966970 & -0.1733140474651558 \\
          1 &    0.65867088851932470219 &  0.7915823554184969 \\
          2 &    2.51962151763406920996 &  2.501600214006679 \\
          3 &    4.63865676589239143594 &  4.640576944608262 \\
         ...&    ...                    &  ...                \\
        100 &    619.636315269976695643 &  619.6363152899056  \\
        101 &    628.028868586785663262 &  628.0288686061785  \\
        103 &    644.899655331227155754 &  644.8996553496056  \\
        105 &    661.883432587416475526 &  661.8834326048518  \\
         ...&    ...                    &  ...                \\
        495 &    5329.38195574207411204 &  5329.381955742333 \\
        496 &    5343.81915399412485822 &  5343.819153994382 \\
        497 &    5358.26617031658854323 &  5358.266170316844 \\
        500 &    5401.66599550160242755 &  5401.665995501854 \\

\hline
\end{tabular}
\end{center}
\label{tab2}
\end{table}

\end{document}